\documentclass[conference]{IEEEtran}
\usepackage{comment}
\usepackage{subcaption}

\usepackage{cite}
\usepackage{amsmath,amssymb,amsfonts}
\usepackage{graphicx}
\usepackage{textcomp}

\usepackage{cite}
\usepackage{amsmath,amssymb,amsfonts}

\usepackage{graphicx}
\usepackage{epstopdf}
 \graphicspath{{\figures}}
\usepackage{textcomp}
\usepackage{xcolor}
\usepackage{verbatim}
\usepackage{makecell}
\usepackage{booktabs}
\usepackage{CJK}
\usepackage{multirow}
\usepackage{balance}

\usepackage{algorithm}
\usepackage{algpseudocode}

\makeatletter
\newcommand{\removelatexerror}{\let\@latex@error\@gobble}
\makeatother

\IEEEoverridecommandlockouts   % show \thanks
\begin{document}

% \begin{CJK}{UTF8}{gbsn}

\title{
% \huge
%Image Semantic Communication Based on Text-to-Image Generative Models 
% Conditional Diffusion-Enhanced Generative Image Semantic Communication with Latent Feature Guidance
% High-Fidelity Generative Image SemCom: Latent Feature guided Conditional Diffusion
Latent Feature-Guided Conditional Diffusion for Generative Image Semantic Communication
% Conditional Diffusion-Enhanced High-Fidelity Generative Image Semantic Communication with Latent Feature Guidance
% Efficient Generative Image Semantic Communication with Conditional Diffusion Enabled Latent Feature Transmission
}
\author{\IEEEauthorblockN{
Zehao Chen\IEEEauthorrefmark{1},
Xinfeng Wei\IEEEauthorrefmark{1},
Haonan Tong\IEEEauthorrefmark{1}, 
Zhaohui Yang\IEEEauthorrefmark{2},
and Changchuan Yin\IEEEauthorrefmark{1}}
% \vspace{0.3cm}

\small \IEEEauthorrefmark{1}
Beijing Laboratory of Advanced Information Network, Beijing University of Posts and Telecommunications, Beijing, China\\
\IEEEauthorrefmark{2} College of Information Science and Electronic Engineering, Zhejiang University, Hangzhou, China\\
Emails: \{chenzhzhz, xinfengwei, hntong,  and ccyin\}@bupt.edu.cn, yang\_zhaohui@zju.edu.cn

% \thanks{This work was supported in part by Beijing Natural Science Foundation under Grant L223027, the National Natural Science Foundation of China under Grants 62471056, 61629101 and 61671086, the 111 Project under Grant B17007.}
}

% make the title area
\maketitle
\vspace{-0.8cm}

\pagestyle{empty}  % no page number for the second and the later pages
\thispagestyle{empty} % no page number for the first page
\begin{abstract}
% Semantic communication is proposed and expected to improve the efficiency and effectiveness of massive data transmission using increasing computing capacity on transceivers. However, existing deep learning-based joint source and channel coding (DeepJSCC) predominantly focus on optimizing pixel-level metrics, and often neglect human perceptual requirements and result in degraded perceptual quality. To address this issue in DeepJSCC methods, we propose a latent representation-oriented image semantic communication system (LRISC). The LRISC transmits image latent semantic features for image generation with semantic consistency, thereby ensuring the perceptual quality at the receiver. In particular, we first map a source image to latent features in a high-dimensional semantic space via a neural network~(NN)-based non-linear transformation. Subsequently, these features are encoded using a joint source-channel coding (JSCC) scheme with adaptive coding length for efficient transmission over a wireless channel. At the receiver, a conditional diffusion model employs the received latent features as conditional guidance to steer the reverse diffusion process, progressively reconstructing high-fidelity images while preserving semantic consistency. Experiments show that the proposed method significantly outperforms existing methods, in terms of learned perceptual image patch similarity and robustness against channel noise,  with an average learned perceptual image patch similarity (LPIPS) reduction of 43.3\% compared to DeepJSCC, while guaranteeing the semantic consistency.

Semantic communication is proposed and expected to improve the efficiency of massive data transmission over sixth generation (6G) networks.
However, existing image semantic communication schemes are primarily focused on optimizing pixel-level metrics, while neglecting the crucial aspect of region of interest (ROI) preservation. 
% However, existing deep learning-based joint source and channel coding (DeepJSCC) image semantic communication scheme predominantly focuses on optimizing pixel-level metrics, while neglecting the crucial aspect of region of interest (ROI) preservation. 
To address this issue, we propose an ROI-aware latent representation-oriented image semantic communication (LRISC) system.
% Specifically,  we first maps a source image to latent features in a high-dimensional semantic space through a neural network based non-linear transformation, followed by joint source-channel coding scheme for transmission through wireless channel. 
In particular, we first map the source image to latent features in a high-dimensional semantic space, these latent features are then fused with ROI mask through a feature-weighting mechanism.
Subsequently, these features are encoded using a joint source and channel coding (JSCC) scheme with adaptive rate for efficient transmission over a wireless channel.
% At the receiver, a conditional diffusion model utilizes the received latent features as conditional information to guide the diffusion process and gradually reconstruct high-quality images with semantic consistency. 
At the receiver, a conditional diffusion model is developed by using the received latent features as conditional guidance to steer the reverse diffusion process, progressively reconstructing high-fidelity images while preserving semantic consistency.
% The LRISC system optimizes both communication efficiency and perceptual accuracy by preserving semantic integrity while enhancing perceptual quality. 
% The proposed LRISC jointly optimizes communication efficiency and perceptual accuracy, ensuring robust semantic integrity while maximizing perceptual quality in the reconstructed output.
Moreover, we introduce a channel signal-to-noise ratio (SNR) adaptation mechanism, allowing one model to work across various channel states.
Experiments show that the proposed method significantly outperforms existing methods, in terms of learned perceptual image patch similarity (LPIPS) and robustness against channel noise,  with an average LPIPS reduction of 43.3\% compared to DeepJSCC, while guaranteeing the semantic consistency.

\end{abstract}


\IEEEpeerreviewmaketitle
% \vspace{-1.5em}

% \vspace{-0.5cm}
\section{Introduction}
With the development of sixth generation (6G) networks, the demands for massive connectivity and low-latency data transmission are rapidly rising\cite{zhang2022toward}. 
Traditional bit stream-based methods struggle with massive data and the increasing number of device connections\cite{barbarossa2023semantic,dai2022communication,10448235}. 
To address this issue, semantic communication, which focuses on conveying the underlying meaning rather than raw data, has emerged as a promising solution to reduce transmission redundancy and adapt to dynamic network environments.
Generative models, combined with joint source-channel coding (JSCC), 
improve transmission efficiency and robustness with enabling semantic extraction and coding process. 
% enables efficient semantic communication by enhancing semantic extraction and coding process, improving transmission efficiency and robustness in wireless channel. 
% Although existing generative semantic communication can significantly improve communication efficiency, the prior arts still lack designs to ensure semantic consistency and adaptability in  generative image semantic communication.
By effectively capturing the characteristics of data distribution, generative models can significantly mitigate the semantic distortion in deep joint source-channel coding (DeepJSCC)\cite{erdemir2023generative}, demonstrating breakthrough potential in the field of generative semantic communication\cite{erdemir2023generative,wu2024cddm,qiao2024latency}. 
The works in \cite{qiao2024latency} first achieved semantic compression by refining images by key semantic features, using semantic features as guiding signals during the decoding process. 
Furthermore, to enhance the perceptual similarity of reconstruction, the study in \cite{lei2023text+} extracted multimodal features and combined them with channel state information as conditions to guide the diffusion model in gradually transforming initial random noise into the final reconstructed data. 
To realize rate adaptive control, \cite{yang2024rate} proposed a generative DeepJSCC approach, utilizing entropy models to estimate entropy of transmitted symbols, and use these symbols to guide the diffusion process for image reconstruction. \cite{wang2025diffcom} proposed a novel generative communication system that leverages diffusion models and stochastic posterior sampling to enhance perceptual quality and robustness.
Despite the improvements achieved by methods in~\cite{erdemir2023generative,wu2024cddm,qiao2024latency,lei2023text+,yang2024diffusion}, they typically pursue average optimality in overall reconstruction quality, neglecting certain regions that are particularly critical for tasks or human perception—known as Regions of Interest (ROI). These regions carry the most semantically valuable information in an image, and accurately restoring the semantic content of ROI is more important than reconstructing fine details across the entire image. Moreover, these methods typically train separate models for discrete SNR levels, lacking explicit mechanisms to adapt to varying channel conditions - a critical limitation for practical deployment.

% they typically treat the pre-trained diffusion models as independent modules, which is not ideal for image transmission tasks that require consistency between the generated and origial source images. 
% Additionally, these methods often require extra data transmission to guide the reverse diffusion process, increasing system overheads. 
% Furthermore, existing methods usually perform fixed data rate, unable to adaptively determine the optimal coding rate for diverse data source. 

\begin{figure*}[t]
    \vspace{-0.3cm}
    \centering
    \includegraphics[width=0.75\textwidth]{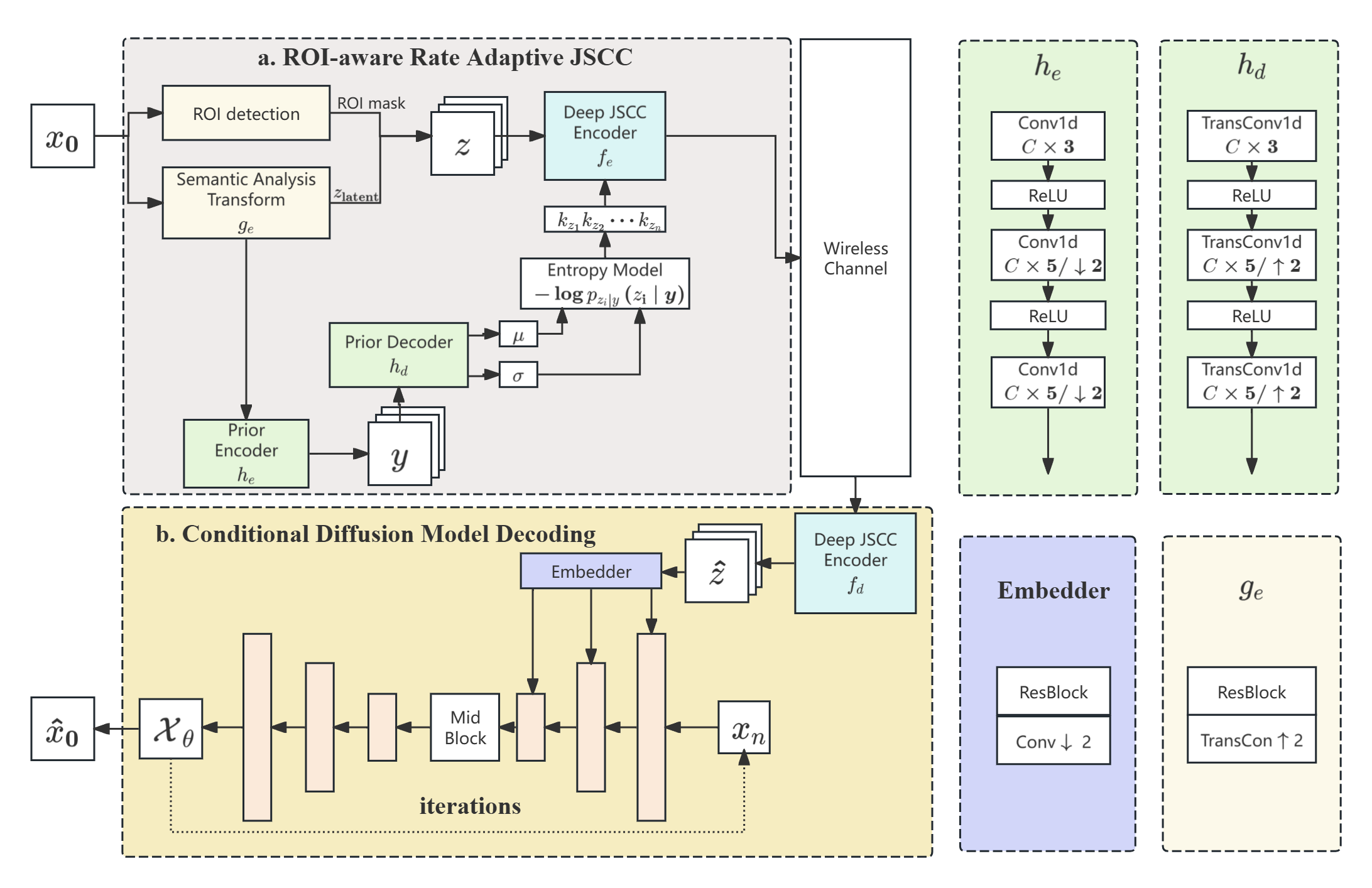}
    % {architecture.png}
    \caption[The framework of proposed LRISC architecture.]{\centering The framework of semantic communication for networks.}
    \label{fig1}
    \vspace{-0.3cm}
\end{figure*}

Inspired by \cite{yang2024rate}, we propose a generative semantic communication system that incorporates adaptive ROI enhancement for image reconstruction.
The key contributions of this work are:
% \begin{itemize}
    % \item 
    (1) We propose a ROI-aware generative image semantic communication framework. The method maps the original image into a latent feature, thereby effectively reducing the amount of data transmitted. Additionally, an ROI importance weighting mechanism is introduced to enhance specific regions of the latent representation, enabling prioritized representation and reconstruction of key semantic areas. 
    (2) To improve robustness under varying channel states, we introduce an SNR-adaptive module that dynamically adjusts the deep joint source-channel coding (JSCC) encoder-decoder’s intermediate features based on real-time channel state information (CSI). This enables a single model to adapt seamlessly across different signal-to-noise ratio (SNR) conditions without requiring multiple trained instances.
    (3) We design a regionally optimized loss function that employs distinct optimization strategies for different image areas, mean squared error (MSE) loss is applied to key regions to preserve structural integrity and a combined MSE and perceptual loss enhances visual quality in non-critical regions, balancing detail fidelity and natural appearance.
    % To train the codecs in the LRISC framework, a perceptual distortion term is incorporated into the distortion loss function to improve the global subjective perceptual quality of the reconstructed images at the receiver. 
Simulation results demonstrate that the proposed method outperforms existing approaches under limited bandwidth scenarios, achieving better semantic consistency and perceptual quality.

The rest of this paper is organized as follows. The system model is introduced in Section \uppercase\expandafter{\romannumeral2}.  The model architecture and training strategy is detailed in Section \uppercase\expandafter{\romannumeral3}. Section \uppercase\expandafter{\romannumeral4} presents the simulation results and analysis to demonstrate the effectiveness of proposed system. Finally, Section \uppercase\expandafter{\romannumeral5} concludes the paper.

% By guiding the diffusion process with latent features, we ensure consistency between the reconstructed and original images. Secondly, we introduce variational entropy modeling to estimate the entropy distribution of latent features, thereby guiding the rate allocation of source-channel coding to maximize the system's coding gain. 

% This paper proposes a novel image semantic transmission framework based on nonlinear transform and a conditional diffusion model, enabling adaptive rate allocation for latent features and guiding the conditional diffusion model with latent features to achieve high perceptual quality image  reconstruction.

% Additionally, a perceptual distortion term is incorporated into the rate-distortion loss function to improve the global subjective perceptual quality of the reconstructed images at the receiver. 

% Finally, simulation results demonstrate that the proposed method outperforms existing approaches under limited bandwidth scenarios, achieving better semantic consistency and perceptual quality.

\section{System Model}
   % In LRISC system, we first map the source image $\boldsymbol{x}$ into a latent feature $\boldsymbol{z}$ through a non-linear transformation, followed by JSCC \cite{dai2022nonlinear}. At receiver, a conditional diffusion model is introduced as the decoder\cite{yang2023lossy}, which utilizes the received latent features $\hat{\boldsymbol{z}}$ as conditional information to guide the denoising reverse diffusion process and iteratively reconstruct images.

   LRISC system integrates an ROI-aware rate-adaptive JSCC with a conditional diffusion model \cite{yang2024rate}. The latent feature $\boldsymbol{z}$ is first obtained through adaptive weighting between the ROI mask and input latent feature $\boldsymbol{z}_\text{latent}$, which is then transmitted via the rate-adaptive JSCC that dynamically optimizes bandwidth allocation\cite{dai2022nonlinear}. Finally, the received $\boldsymbol{z}$ serves as conditional guidance for the diffusion model to generate high-fidelity reconstructions \cite{yang2023lossy}. This unified architecture achieves enhanced reconstruction quality through coordinated ROI-aware latent feature encoding, rate-adaptive channel transmission, and semantic-guided diffusion model decoding.

\begin{figure*}[t]
\centering
\setlength{\belowcaptionskip}{-0.45cm}
\includegraphics[width=0.75\textwidth]{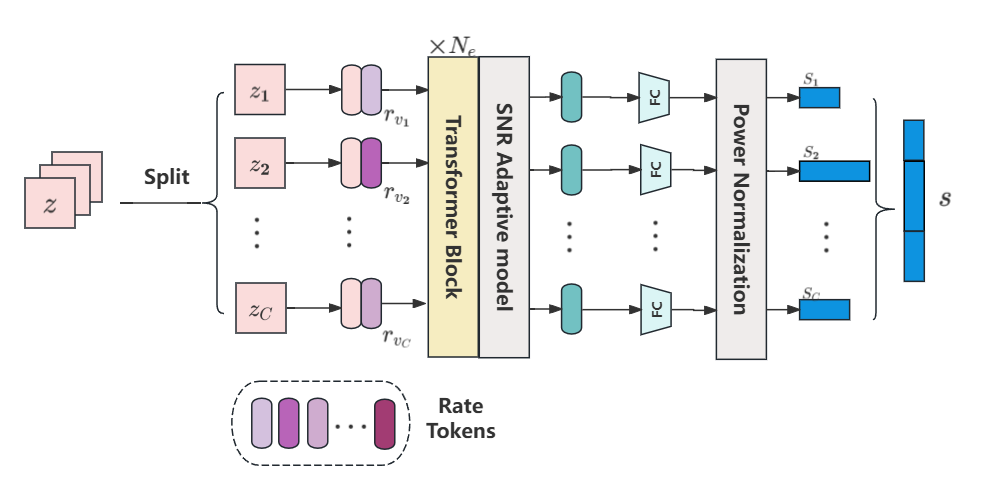}
% {JSCCEncoder.png}
\centering
\vspace{-0.4cm}
\caption{The architecture of JSCC encoder. 
%The structure of DeepJSCC decoder is with a mirrored architecture.
}
\label{fig2}
\vspace{-0.4cm}
\end{figure*}

\subsection{ROI-aware latent feature encoding}
As shown in Fig. \ref{fig1}, at the transmitter, a nonlinear analytical transform $g_e$ is employed to extract the semantic feature in the latent-space from the source image $\boldsymbol{x}\in \mathbb{R}^{n\times n}$, denoted as  $\boldsymbol{z_\text{latent}}=g_e\left({\boldsymbol x} ; {\boldsymbol \theta}_g\right)$. To optimize visual quality, we allocate more bits to the ROI, thereby enhancing the reconstruction accuracy of critical features. 
To distinguish between the focused area and the background, we utilize a saliency detection network $\mathcal{S}$ \cite{chen2020global} to generate an ROI mask $\boldsymbol{m_\text{ROI}}$ offline. A convolutional layer is then applied to smooth the saliency map, mitigating sharp boundaries and resulting in:
\begin{equation}
\boldsymbol{m}_\text{ROI} = \sigma\big(\text{Conv}(\mathcal{S}(\boldsymbol{x}))\big),
\end{equation}
where $\sigma$ denotes the sigmoid function.

Since the spatial characteristics of the image are preserved when mapped into latent features by the encoder, we derive an ROI mask $\boldsymbol{m}_\text{latent}$ for $\boldsymbol{z}_\text{latent}$ via average pooling: 
\begin{equation}
    \boldsymbol{m}_{latent} = AvgPool(\boldsymbol{m}_\text{ROI}).
\end{equation}

By applying the weighted mask $\boldsymbol{m}_\text{latent}$, the ROI features in the latent space are amplified, thereby allocating more bits to the ROI in the resulting code stream. To further control the rate allocation, we introduce a scaling factor $\gamma$, which adjusts the relative importance of the ROI:
\begin{equation}
    \boldsymbol {z}=\frac{\boldsymbol{m}_{\text {latent }}+\gamma}{\gamma} \otimes \boldsymbol{z}_\text {latent},
\end{equation}
Here, a smaller $\gamma$ means more bits are allocated to the ROI area in latents. What’s more, to prevent the background texture from degrading, we preserve a subset of channels (e.g., the first 64 out of 256) without weighting, ensuring sufficient information is retained for background reconstruction:
\begin{equation}
   \boldsymbol {z}=\boldsymbol{z}_\text {latent ch0-ch63} \| \boldsymbol{z}_{ ch64-ch255}.
\end{equation}
The latent feature $\boldsymbol{z}$ consists of 256 channels, the first 64 channels are protected and the remaining channels are adaptively weighted.

\subsection{Rate-adaptive Channel Transmission}

Adaptive rate allocation is achieved by leveraging entropy measurements. To enable precise entropy estimation, a hyperprior entropy model is introduced to refine entropy estimation accuracy. The hyperprior encoder $h_e$ extracts side information $\boldsymbol{y}=h_e\left(\boldsymbol{z} ; \boldsymbol{\theta}_h\right)$ capturing spatial dependencies in $\boldsymbol{z}$. $\boldsymbol{z}$ contains $k$ embedding vectors $\boldsymbol{z}_i$ of dimension $C$. The conditional probability of $\boldsymbol{z}$ given hyperprior $\boldsymbol{y}$ is formally expressed as:
\begin{equation}
\begin{aligned}
& p_{\boldsymbol{z} \mid \boldsymbol{y}}(\boldsymbol{z} \mid \boldsymbol{y})=\prod_i\left(\mathcal{N}\left(\boldsymbol{z}_i \mid \mu_i, \sigma_i\right) * \mathcal{U}\left(-\frac{1}{2}, \frac{1}{2}\right)\right)\left(\boldsymbol{z}_i\right), \\
& \text { with }(\boldsymbol{\mu}, \boldsymbol{\sigma})=h_d\left({\boldsymbol{y}} ; \boldsymbol{\phi}_h\right),
\end{aligned}
\end{equation}
where the hyperprior decoder $h_d\left({\boldsymbol{y}} ; \boldsymbol{\phi}_h\right)$ generates parameters $(\boldsymbol{\mu}, \boldsymbol{\sigma})$ that characterize a convolved Gaussian-uniform distribution for each latent element $\boldsymbol{z}_i$. This formulation accounts for quantization effects through the uniform noise term $\mathcal{U}\left(-\frac{1}{2}, \frac{1}{2}\right)$, while the Gaussian component $\mathcal{N}\left(\boldsymbol{z}_i \mid \mu_i, \sigma_i\right) $ enables entropy-adaptive coding. 

The channel bandwidth cost $K_z$ for transmitting $\boldsymbol z$ can be given by
\begin{equation}
    K_z=\sum_{i=1}^k \bar{k}_{\boldsymbol{z}_i}=\sum_{i=1}^k Q\left(k_{\boldsymbol{z}_i}\right)=\sum_{i=1}^k Q\left(-\beta \log p_{z_i \mid \boldsymbol{y}}\left(\boldsymbol{z}_i \mid \boldsymbol{y}\right)\right),
\end{equation}
where $\beta$ is a hyperparameter to balance channel bandwidth cost from the estimated entropy, $\bar{k}_{\boldsymbol{z}_i}$ is the bandwidth consumption of $\boldsymbol{z}_i$, $Q$ denotes a $2^n$-level scalar quantization with the quantized value set as $\mathcal{V}=\left\{v_1, v_2, \ldots, v_{2^n}\right\}$.
The process by which JSCC maps the latent features $\boldsymbol z$ into channel transmission symbol $\boldsymbol{s}\in \mathbb{C}^k$ is given by 
\begin{equation}
    \boldsymbol{s}=f_e\left(\boldsymbol{z}, p_{\boldsymbol{z}_i \mid {\boldsymbol{y}}}\left({\boldsymbol{z}}_i \mid {\boldsymbol{y}}\right) ; \boldsymbol{\theta_f}\right).
\end{equation}

 To meet the energy constraints of practical communication systems, the channel transmission symbols must satisfy an average power constraint before transmission, given by 
  $ \frac{1}{k} \sum_{i=1}^k\left|\boldsymbol{s}_i\right|^2 \leq P $
where $P$ is the  maximal transmision power.
Given the channel transmission function $W$, the received symbol sequence is $\hat{\boldsymbol s}=W(\boldsymbol s)$.

% In this paper, we consider AWGN channel, such that the received symbols $\hat{\boldsymbol{s}}$ at the receiver can be given by
% \begin{equation}
%     \hat{\boldsymbol{s}}=W(\boldsymbol{s} \mid {h})={h} * \boldsymbol{s}+\boldsymbol{n},
% \end{equation}
% where $h$ indicates the channel gain, and $\boldsymbol{n} \sim \mathcal{N} \left(0, \sigma_n^2 \ \boldsymbol{I}\right)$ is the additive white Gaussian noise with variance $\sigma_n^2$ and $\boldsymbol{I}$ being identity matrix.The reconstructed latent features $\hat{\boldsymbol{z}}=f_d\left(\hat{\boldsymbol{s}} ; \boldsymbol{\phi}_f\right)$ is obtained through the JSCC decoder $f_d$, where $\boldsymbol{\phi}_f$ represents the trainable parameters of the JSCC decoder. A conditional diffusion model is then utilized as the decoder $g_d$ for image reconstruction, resulting in $\hat{\boldsymbol{x}}_0=g_d\left(\hat{\boldsymbol{x}}_N, \hat{\boldsymbol{z}} ; \boldsymbol{\phi}_g\right)$, where $\hat{\boldsymbol{x}}_n$ is randomly sampled Gaussian source.

\subsection{Semantic-guided Diffusion Model Decoding}
At the receiver, we adopt a conditional diffusion model for image reconstruction, received latent feature $\hat{\boldsymbol{z}} = f_d(\hat{\boldsymbol{s}};\boldsymbol{\phi}_f)$ serves as the condition for the reverse process. The diffusion process (denoeted by $q$) gradually adds Gaussian noise to an image $\boldsymbol{x}_0$ over a series of time steps, resulting in a sequence of noisy data $\boldsymbol{x}_1, \boldsymbol{x}_2, \ldots, \boldsymbol{x}_N$. The reverse diffusion process (denoted by $q_\theta$) learns to denoise the data step by step. At step $n$, the two Markov processes $q$ and $p_\theta$ can be respectively described as
\begin{equation}
\label{eq:forward}
    q\left(\boldsymbol{x}_n \mid \boldsymbol{x}_{n-1}\right)=\mathcal{N}\left(\boldsymbol{x}_n | \sqrt{1-\beta_n} \boldsymbol{x}_{n-1}, \beta_n \boldsymbol{I}\right),
\end{equation}
% \begin{equation}
%     p_\theta\left(\boldsymbol{x}_{n-1} \mid \boldsymbol{x}_n\right)=\mathcal{N}\left(\boldsymbol{x}_{n-1} \mid M_\theta\left(\boldsymbol{x}_n, n\right), \beta_n \boldsymbol{I}\right),
% \end{equation}
\begin{equation}
    p_\theta\left(\boldsymbol{x}_{n-1} \mid \boldsymbol{x}_n, \hat{\boldsymbol{z}}\right)=\mathcal{N}\left(\boldsymbol{x}_{n-1} \mid M_\theta\left(\boldsymbol{x}_n, \hat{\boldsymbol{z}}, n\right), \beta_n \boldsymbol{I}\right).
\end{equation}
where variable $\beta_n $ serves as a constant hyperparameter. The reverse process is parameterized by a neural network (NN) $M_\theta\left(\boldsymbol{x}_n, \hat{\boldsymbol{z}}, n\right)$. We use a pixel-space prediction NN $\mathcal{X}_\theta$ to learn to directly  reconstruct image  $x_0$ instead of noise $\epsilon$ \cite{salimans2022progressive}. The loss function can be described as follows
% \begin{equation}
% \label{Eq_noise}
%     \mathcal{L}\left(\theta, x_0\right)=\mathbb{E}_{\boldsymbol{x}_0, n, \epsilon}\left\|\epsilon-\epsilon_\theta\left(\boldsymbol{x}_n, n\right)\right\|^2,
% \end{equation}
\begin{equation}
    \mathcal{L}\left(\theta, x_0\right) = \mathbb{E}_{\boldsymbol{x}_0, n, \epsilon} \frac{\alpha_n}{1-\alpha_n}\left\|\boldsymbol{x}_0-\mathcal{X}_\theta\left(\boldsymbol{x}_n, \hat{\boldsymbol{z}}, \frac{n}{N_{\text {train }}}\right)\right\|^2,
\end{equation}
where $n \sim \operatorname{Unif}\{1, \ldots, N\}$, with $\operatorname{Unif}$ denoting a uniform distribution over the set ${1, 2, \ldots, N}$, $\epsilon \sim \mathcal{N}(\boldsymbol{0}, \boldsymbol{I})$ and $\alpha_n=\prod_{i=1}^n\left(1-\beta_i\right)$.We condition the model on the pseudo-continuous variable $\frac{n}{N_{\text {train}}}$ which offers additional flexibility in choosing the number of denoising steps for decoding. 

Once the model is trained, image can be generated by the denoising diffusion implicit model (DDIM) \cite{song2020denoising} that follows a deterministic generation, which is 
\begin{equation}
    \boldsymbol{x}_{n-1}=\sqrt{\alpha_{n-1}} \mathcal{X}_\theta\left(\boldsymbol{x}_n, \hat{\boldsymbol{z}}, \frac{n}{N}\right)+\sqrt{1-\alpha_{n-1}} \epsilon_\theta\left(\boldsymbol{x}_n, \hat{\boldsymbol{z}}, \frac{n}{N}\right).
\end{equation}

\section{ Model Architecture and Training Strategy  }
This section presents three parts: the NN architecture of LRISC, the system's loss function, and the multi-stage training algorithm.

\begin{figure}[t]
\centering
\includegraphics[width=0.5\textwidth]{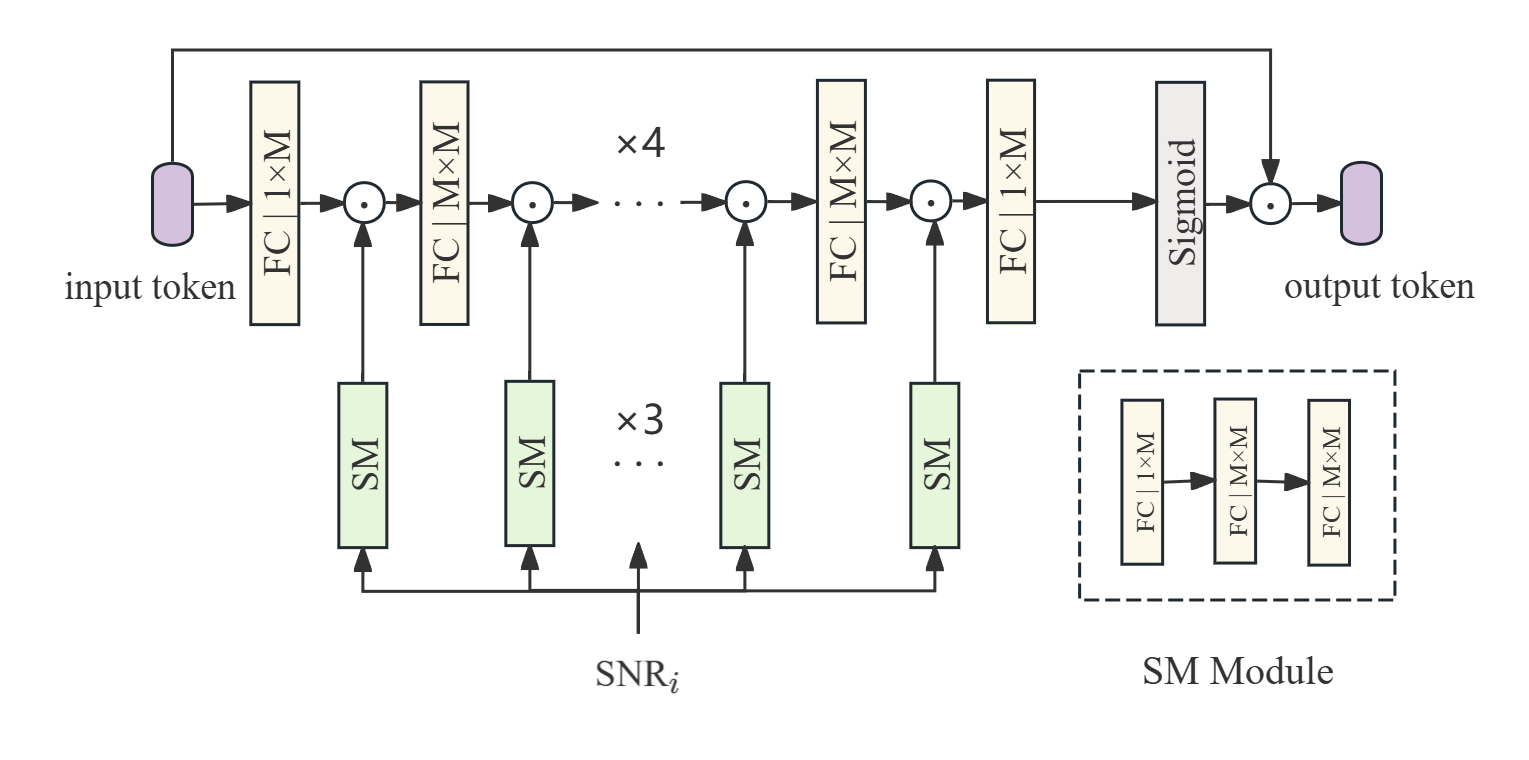}
% {JSCCEncoder.png}
\centering
\vspace{-0.4cm}
\setlength{\abovecaptionskip}{-0.45cm}
\setlength{\belowcaptionskip}{-0.45cm}
\caption{The architecture of SNR adaptive model. FC is fully-connected network, ``$\odot$" is the element-wise product.
%The structure of DeepJSCC decoder is with a mirrored architecture.
}
\label{fig:snr_}
\vspace{-0.4cm}
\end{figure}

\subsection{Model Architecture}
\begin{enumerate}
    \item \textit{\textbf{Prior Codec}} $h_e$, $h_d$: The structure of $h_e$ and $h_d$ is shown in Fig. \ref{fig1}. It consists of fully convolutional layers followed by the ReLU activation functions.
    \item \textit{\textbf{JSCC Codec}} $f_e$, $f_d$: We adopt a pair consisting of a transformer-like %\cite{vaswani2017attention} 
    JSCC encoder and decoder as $f_e$, $f_d$. The encoder consists of $N_e$ transformer blocks and fully connected~(FC) layers to implement rate allocation, as shown in Fig.~\ref{fig2}. Specifically, $\boldsymbol{z}$ is firstly separated into patch embedding sequence ${\boldsymbol{z}_1, \boldsymbol{z}_2, \cdots, \boldsymbol{z}_c}$. Guided by the entropy model $-\log P_{\boldsymbol{z}_i \mid \boldsymbol{y}}\left(\boldsymbol{z}_i \mid \boldsymbol{y}\right)$, a set of learnable rate token embeddings with the same dimension as $\boldsymbol{z}_i$ are developed \cite{dai2022nonlinear}, each of whcih corresponds to a value in $\mathcal{V}=\left\{v_1, v_2, \ldots, v_{2^n}\right\}$, each $\boldsymbol{z}_i$ is merged with its corresponding rate token $r_{vi}$. FC layers  with output dimensions of $v_q$, $q = 1,2,\dots,2^n$ are employed to map the embeddings into $s_i$ with given dimensions.

 \item \textit{\textbf{SNR Adaptive Model}}: The SNR adaptive model is inserted to the last layer of JSCC encoder and the first layer of JSCC decoder, as shown in  Fig~\ref{fig:snr_}. It consists of 8 FC layers alternating with 7 SNR modulation~(SM) modules~\cite{swinJSCC}. SM module is a three-layered FC network, which takes the SNR for received $\boldsymbol{z}_i$~(denoted by  SNR$_i$) as input and transforms it into an $c$-dimensional vector $sm_i$. Multiple SM modules are cascaded sequentially in a coarse-to-fine manner, the previous modulated features are fed into subsequent SM modules.
    \item \textit{\textbf{Conditional Diffusion Model}}: The denoising module utilizes a U-Net structure \cite{ronneberger2015unet,zeng}. Each U-Net unit is composed of two ResNet blocks, an attention block, and a convolutional upsampling and downsampling block. The conditional diffusion model network employs six U-Net units in both the downsampling and upsampling paths. During the downsampling phase, the channel dimension expands according to the formula $64\times j$, where $j$ denotes the layer index ($j=1,2,...,6$). The upsampling units follow the reverse order. For embedding condition $\boldsymbol{z}$, we employ ResNet blocks and transpose convolutions to upscale $\boldsymbol{z}$ to match the inputs' dimensions of the initial four U-Net downsampling units.
\end{enumerate}

\subsection{Loss Function}
The optimization objective is to achieve a trade-off between compression performance and image reconstruction quality.
To enhance the focus on ROI area while maintaining the perceptual quality of the overall image, a regionally differentiated distortion function is designed under the guidance of an ROI mask to optimize the reconstruction quality of both ROI area and the background area. Specifically, separate distortion losses were defined for the ROI region $d_{\mathrm{ROI}}$ and the background region $d_{\mathrm{BG}}$. The ROI region emphasizes pixel-level reconstruction accuracy in critical semantic areas, as shown in below:
\begin{equation}
    d_{\mathrm{ROI}}(x, \hat{x})=\boldsymbol{m}_{2 \mathrm{D}} \otimes \operatorname{MSE}(x, \hat{x}),
\end{equation}
MSE distortion measures the pixel-level differences between a reconstructed image and the original image, ensuring image fidelity.
For background regions, a combination of MSE and perceptual loss terms is adopted to enhance the overall perceptual quality of the image, ensuring that the reconstructed image aligns more closely with human visual perception in terms of structure and texture, resulting in:
\begin{equation}
    d_{\mathrm{BG}}(x, \hat{x})=\operatorname{MSE}(x, \hat{x})+\eta \cdot d_P(x, \hat{x}),
\end{equation}
Here, $d_p(\cdot, \cdot)$ represents the perceptual loss term, which employs LPIPS as the perceptual metric. It is used to balance the trade-off between MSE distortion and perceptual loss. The overall distortion $d\left(\boldsymbol{x}_0, \hat{\boldsymbol{x}}_0\right)$ of the reconstructed image can be expressed as:
\begin{equation}
    \begin{aligned}
    \label{loss}
L & = D + \lambda \cdot R \\
& = \mathbb{E}_{\boldsymbol{x}_0} \bigl[ -\lambda \log p_{\boldsymbol{z} \mid \boldsymbol{y}}(\boldsymbol{z} \mid \boldsymbol{y}) 
   - \lambda \log p_{\boldsymbol{y}}(\boldsymbol{y}) \\
& \quad + d\left(\boldsymbol{x}_0, \hat{\boldsymbol{x}}_0\right) 
   + d\left(\boldsymbol{x}_0, \overline{\boldsymbol{x}}_0\right) \bigr],
\end{aligned}
\end{equation}
Here, $\lambda$ denotes the hyperparameter that controls the trade-off between rate and distortion, R represents the total channel transmission rate, and D indicates the end-to-end distortion. $d\left(\boldsymbol{x}_0, \hat{\boldsymbol{x}}_0\right)$ measures the distortion between the original image and the reconstructed image, while $d\left(\boldsymbol{x}_0, \overline{\boldsymbol{x}}_0\right)$ measures the distortion between the original image and the compressed image.

\subsection{Training Strategy}

\begin{algorithm}[tb] % [tb] 表示浮动位置为顶部或底部
\small
\caption{Training procedure for the proposed LRISC} % 算法标题
\label{alg:training} % 标签用于引用
\begin{algorithmic}[1] % 每行显示行号
    \renewcommand{\algorithmicrequire}{\textbf{Input:}}
    \Require Training dataset $X$, the Lagrange multiplier $\lambda$ on the rate term, perception and distortion trade-off parameter $\eta$, scaling factor $\beta$ from entropy to channel bandwidth cost and learning rate $l_r$.
    
    \State \textbf{Stage 1: Train $g_e,$ $g_d,$ $h_e,$ 
   $h_d$}
    \State Randomly initialize all parameters and freeze the parameters of $f_e$ and $f_d$
    
   %  \For{each epoch} % 第一阶段循环
   %      \State Sample $\boldsymbol{x} \sim p_{\boldsymbol{X}}$
   %      \State Calculate the loss function: 
   %      \State $\begin{aligned}  \mathcal{L_{RDP}} = \mathbb{E}_{\boldsymbol{x}_0}[
   %  (1-\eta)d(\boldsymbol{x}_0, \overline{\boldsymbol{x}}_0)] 
   % + \eta[d_p(\boldsymbol{x}_0, \overline{\boldsymbol{x}}_0)] \\
   % + \lambda[-\log p_{{\boldsymbol{z}}\mid {y}}({\boldsymbol{z}} \mid {\boldsymbol{y}})-\log p_{{y}}({\boldsymbol{y}})]\end{aligned}$
   %      \State Update the parameters $\left(\boldsymbol{\theta}_g, \boldsymbol{\phi}_g, \boldsymbol{\theta}_h, \boldsymbol{\phi}_h\right)$
   %  \EndFor
   
    \For{training iteration $1$ to $N_{\text {train }}$}
        \State sample $\boldsymbol{x} \sim p_{\boldsymbol{X}}$, $n \sim \mathcal{U}\left(0,1,2, . ., N_{\text {train }}\right)$, $\epsilon \sim \mathcal{N}(\mathbf{0}, \mathbf{I})$
        \State ${\overline{\boldsymbol{x}}}_n=\sqrt{\alpha_n} \boldsymbol{x}_0+\sqrt{1-\alpha_n}\epsilon)$
        %\State ${\boldsymbol{z}}=g_e\left(\boldsymbol{x}_0\right)+\mathcal{U}(-0.5,0.5)$
        \State $\overline{\boldsymbol{x}}_0=\mathcal{X}_\theta\left(\overline{\boldsymbol{x}}_n, n / N_{\text {train }}, {\boldsymbol{z}}\right)$
        \State $\begin{aligned}  \mathcal{L} = \mathbb{E}_{\boldsymbol{x}_0}[
    d(\boldsymbol{x}_0, \overline{\boldsymbol{x}}_0) 
   + \lambda[-\log p_{{\boldsymbol{z}}\mid {y}}({\boldsymbol{z}} \mid {\boldsymbol{y}})-\log p_{{y}}({\boldsymbol{y}})]\end{aligned}$
        \State Update the parameters $\left(\boldsymbol{\theta}_g, \boldsymbol{\phi}_g, \boldsymbol{\theta}_h, \boldsymbol{\phi}_h\right)$
   \EndFor

    \State \textbf{Stage 2: Train $f_e,$ $f_d$}
    \State Load and freeze the parameters trained in Stage 1 and randomly initialize $f_e$ and $f_d$
    \For{each epoch} % 第二阶段循环
        \State Sample $\boldsymbol{x} \sim p_{\boldsymbol{X}}$
        \State Calculate the loss function based on Eq.(\ref{loss})
        \State Update parameters: $\left(\boldsymbol{\theta}_f, \phi_f\right)$
    \EndFor
    
    \State \textbf{Stage 3: Fine-tune the whole model}
    \State Load the parameters trained in the previous stages
    \For{each epoch} % 第三阶段循环
        \State Repeat steps 14 to 16
        \State Update parameters $\left(\boldsymbol{\theta}_g, \boldsymbol{\theta}_h, \phi_g, \phi_h, \boldsymbol{\theta}_f, \phi_f\right)$
    \EndFor

        \renewcommand{\algorithmicensure}{\textbf{Output:}}
    \Ensure Parameters $\left(\boldsymbol{\theta}_g^*, \phi_g^*, 
    \boldsymbol{\theta}_h^*, \phi_h^*, \boldsymbol{\theta}_f^*, \phi_f^*\right)$

\end{algorithmic}
\end{algorithm}

To ensure training stability and enhance overall performance, we propose a multi-stage training strategy. Initially, we train each module individually~\cite{Xwei} to reduce complexity and facilitate convergence~\cite{W2V_}. Once all modules have been trained separately, we fine-tune the entire model. The complete multi-stage training process is shown in Algorithm 1. First, train $g_e$, $g_d$, $h_e$ and $h_d$. Secondly, train JSCC codecs $f_e$ and $f_d$ with channel data. At last, fine-tune all modules from the aforementioned two steps in an end-to-end optimization manner.

\section{Simulation Analysis and Discussions}
%In this section, we conduct simulations to assess the performance of the proposed LRISC. 

\subsection{Simulation Settings}
For our experiments, we use the COCO dataset %\cite{lin2015microsoftcococommonobjects}%
for training, and we select 50,000 images, which are randomly cropped to 256 × 256, and the Kodak dataset for testing. We implement the proposed LRISC model using the PyTorch framework. We exploit the Adam optimizer with an initial learning rate of $1 \times 10^{-4}$, and the batch size is set to 16. 
In the first stage, we randomly initialize the model parameters and train $g_e$, $g_d$, $h_e$ and $h_d$, the training steps set to $10^{6}$. In the second stage, we first train JSCC codec $f_e$ and $f_d$ with a fixed channel state (SNR = 10dB) except for the snr adaptive module. Then, the whole JSCC codec is trained with snr adaptive module with variable channel state. Finally, during fine-tuning, all model parameters are unfrozen, and the model is trained on the full COCO training dataset for another 20 epochs, with a gradually decaying learning rate.
 The scaling factor $\beta$ is set to 0.2 at SNR = 10dB, the perception and distortion trade-off parameter $\eta$ is set to 0.5. The training process is conducted on an NVIDIA RTX 4090 GPU to accelerate computation. 
%These details provide a comprehensive overview of the experimental setup and ensure reproducibility of the results.

\subsection{Performance Comparison}
We compare our proposed method with two widely recognized baselines: 1) the classical DeepJSCC model optimized using MSE distortion \cite{8683463} and 2) the traditional source-channel separation scheme. Specifically, the source-channel separation scheme uses the BPG for source coding, combined with  Low-Density Parity-Check (LDPC) for channel coding, which is marked as ``BPG + LDPC''. 

For the performance evaluation, we utilize two commonly employed metrics: LPIPS and peak signal-to-niose-ratio (PSNR). LPIPS is used for measuring perceptual similarity, as it correlates more closely with human visual perception than traditional pixel-based metrics like PSNR. 
While PSNR provides a quantitative measure of the reconstruction quality based on pixel differences.

\begin{figure}[t]
\centering
\setlength{\belowcaptionskip}{-0.45cm}
\includegraphics[width=0.5\textwidth]{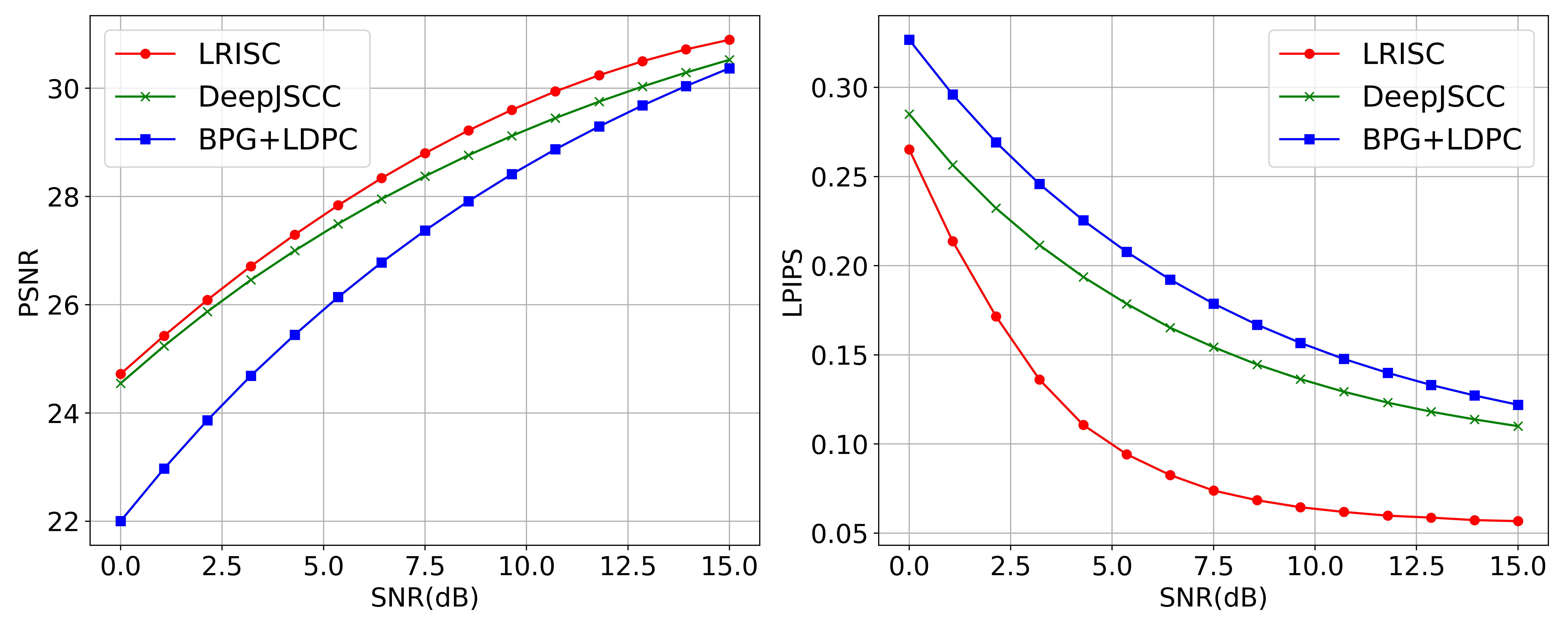}
\centering
\caption{PSNR and LPIPS performance versus SNR.}
\label{fig:snr}
\vspace{-0.45cm}
\end{figure}

\begin{figure*}[t]
\centering
\setlength{\belowcaptionskip}{-0.45cm}
\begin{minipage}{0.32\textwidth}
    \centering
    \includegraphics[width=\linewidth]{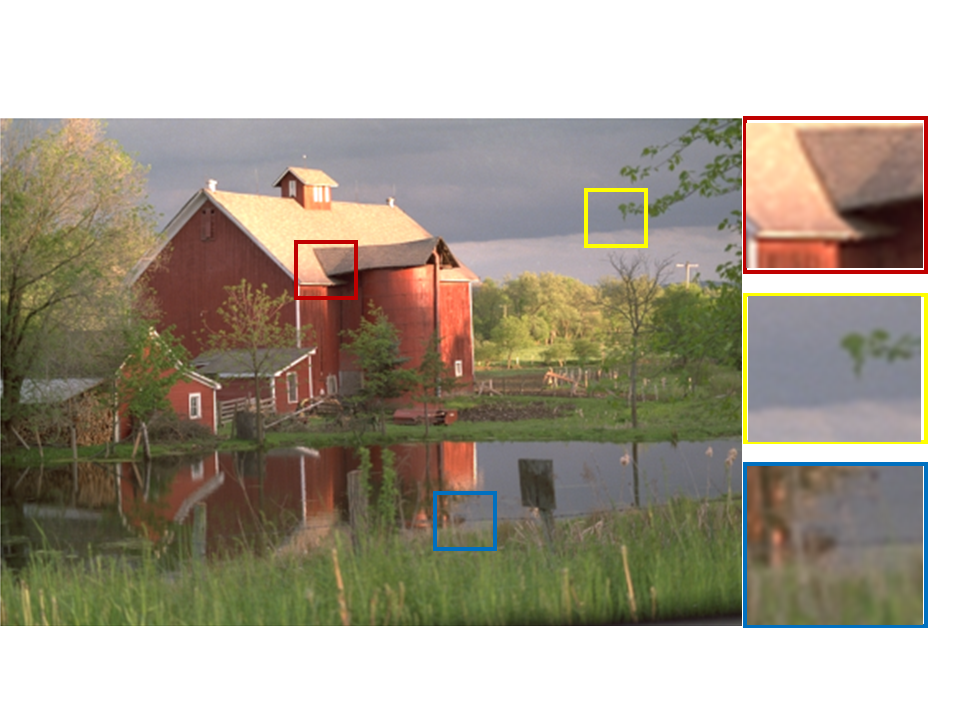}
    \subcaption{ (a) }\label{fig:a}
\end{minipage} \hfill
\begin{minipage}{0.32\textwidth}
    \centering
    \includegraphics[width=\linewidth]{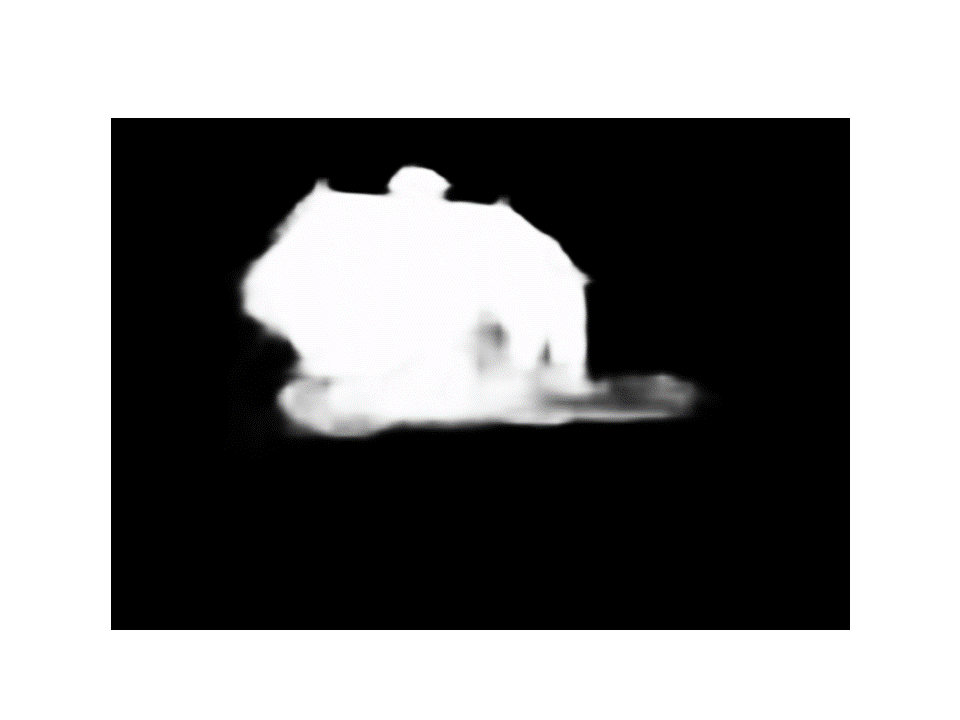}
    \subcaption{ (b) }\label{fig:b}
\end{minipage} \hfill
\begin{minipage}{0.32\textwidth}
    \centering
    \includegraphics[width=\linewidth]{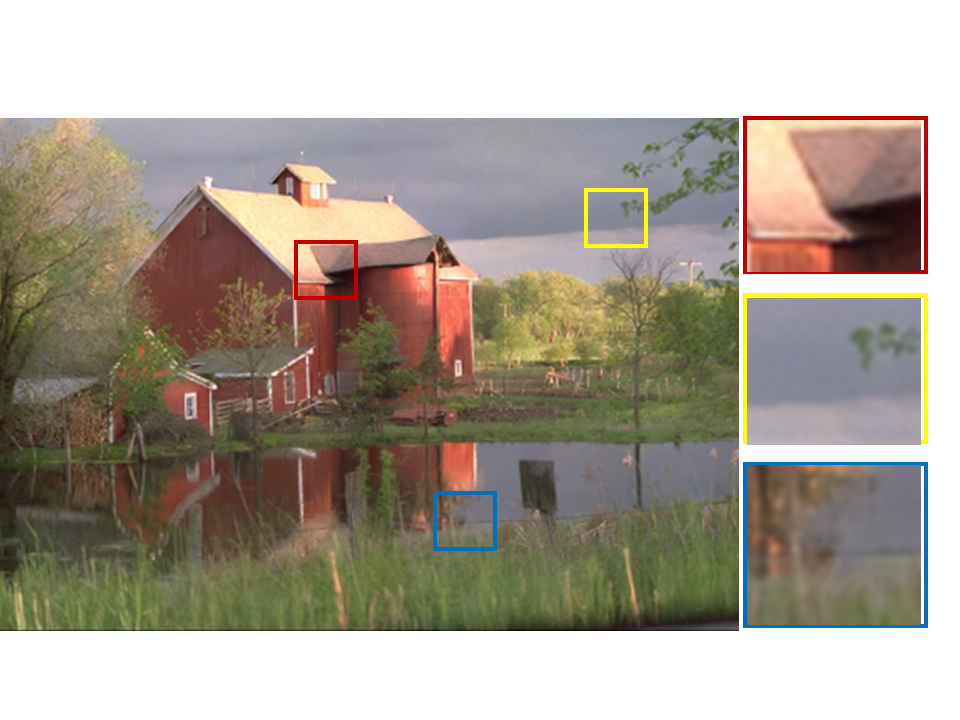}
    \subcaption{ (c) }\label{fig:c}
\end{minipage}
\caption{ (a) Origial image; ~~ (b) The image's ROI area; ~~(c) Reconstructed image with low CBR }\label{fig:visual}
\setlength{\belowcaptionskip}{-0.45cm}
\end{figure*}
\vspace{-0.2cm}

Fig. \ref{fig:snr} compares the performance across varying SNR levels. 
LRISC consistently achieves the highest PSNR across the entire SNR range. Notably, at lower SNR values, LRISC and DeepJSCC demonstrate significant robustness, and achieves a reduction of nearly 43.3\% in LPIPS compared to DeepJSCC.
This improvement highlights the effectiveness of LRISC’s optimized JSCC strategy in preserving image quality under noisy channel conditions. 
When examining LPIPS, the performance gap between LRISC and the baseline methods widens as SNR increases. 
This improvement is because LRISC generates images' perceptual information, with more semantics aligning to the original images. 

\begin{figure}[t]
\centering
\setlength{\belowcaptionskip}{-0.45cm}
\includegraphics[width=0.5\textwidth]{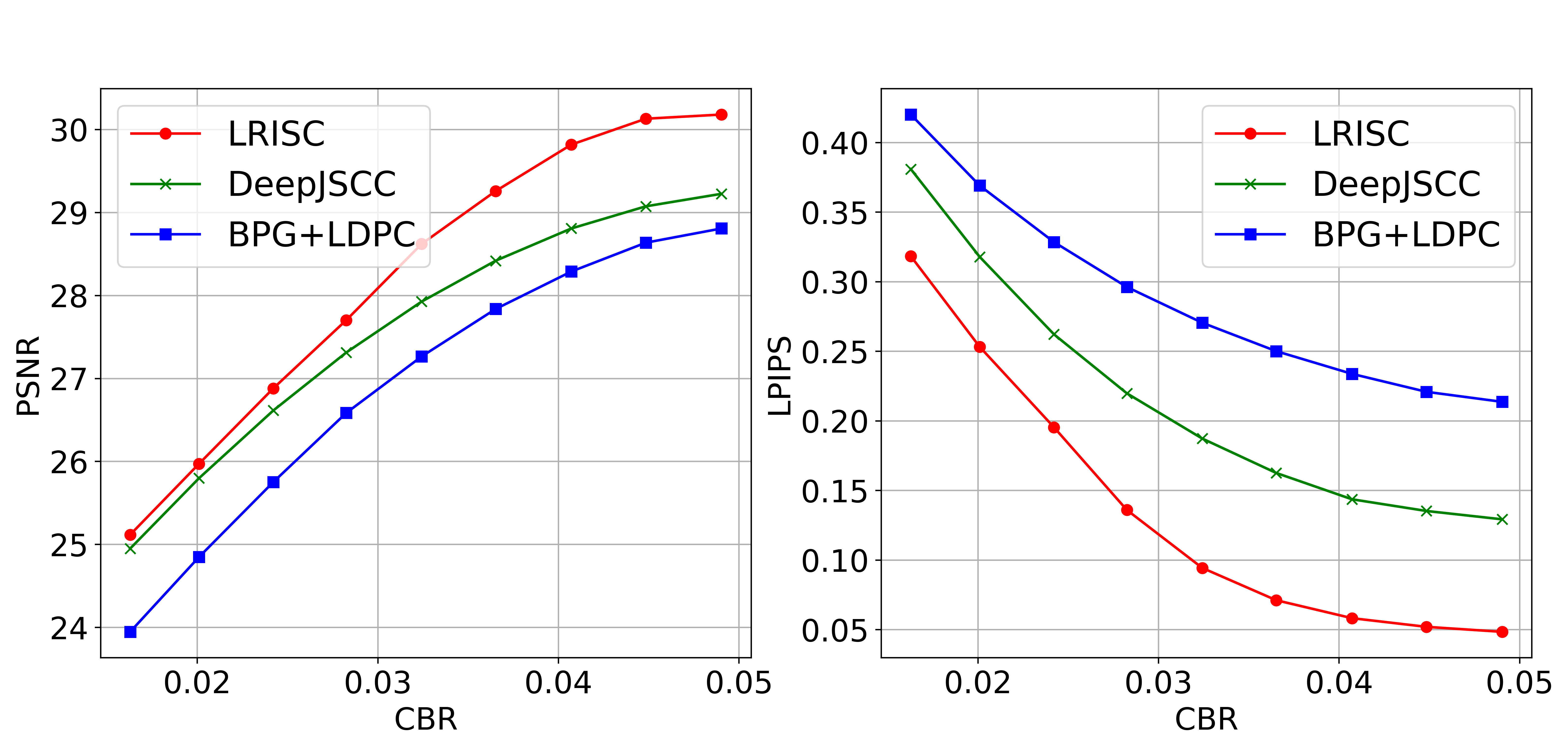}
\centering
\caption{PSNR and LPIPS performance versus CBR.}
\label{fig:cbr}
\vspace{-0.4cm}
\end{figure}

Fig. \ref{fig:cbr} demonstrates the performance comparison for different methods in terms of PSNR as the channel bandwidth ratio (CBR) varies. 
Specifically, LRISC consistently outperforms the DeepJSCC and ``BPG + LDPC'' baselines in terms of PSNR. 
This improvement is attributed to the enhanced compression and adaptive coding in our LRISC, which enables higher reconstruction quality with lower CBR. 
Additionally, LRISC achieves the lowest LPIPS value, indicating superior perceptual fidelity compared to other methods. 
In contrast, the BPG+LDPC baseline exhibits the highest LPIPS values, indicating that while it maintains acceptable pixel-level quality, its perceptual quality remains suboptimal.

% Fig.~\ref{fig:visual} shows visual comparisons of image transmission quality across different SNR values in our proposed framework. 
% From Fig.~\ref{fig:visual} we see that, 
% at the lowest SNR level (i.e., 0 dB), the reconstructed images suffer from severe quality degradation, exhibiting prominent visual distortions, while still avoid ``cliff-effect"~\cite{Pan} 
% in traditional methods. 
% As the SNR increases to 5 dB, these distortions are substantially alleviated. When the SNR reaches 10 dB or higher, the reconstructed image achieves near-identical fidelity to the original source.
% This is because the transmitted latent features effectively capture semantic information thus maintaining robust semantic consistency under low-SNR conditions, and significantly enhancing perceptual quality of the reconstructed images.

Fig.~\ref{fig:visual} shows the image transmission quality under low CBR conditions. Fig.~\ref{fig:visual} (a) shows the transmitted origial image, while Fig.~\ref{fig:visual} (b) presents the ROI map generated by our system. The ROI map highlights the key areas of the image, which are allocated more bits to ensure reconstruction accuracy. In Fig.~\ref{fig:visual} (c), we demonstrate the transmitted image under a low channel bit rate of 0.017 and an SNR of 10. It is clear that, even under low CBR conditions, our system effectively preserves the semantic content of the image, particularly in the ROI region. Despite some minor distortions, the overall quality of the transmitted image is very close to the original. The combination of the ROI-aware mechanism and adaptive rate control enables our system to maintain high perceptual quality with low bit rate conditions, demonstrating robustness in suboptimal transmission environments while ensuring semantic consistency and visual quality.

% \vspace{-0.2cm}

\section{Conclusion}
In this paper, we propose an ROI-aware generative image semantic communication system driven by latent features LRISC. Specifically, the system incorporates an adaptive ROI-guided feature encoding mechanism that intelligently allocates transmission rates, combined with a powerful diffusion model to achieve high-fidelity image reconstruction under limited bandwidth. Simulation results demonstrate that the system not only maintains robustness in semantic image transmission but also delivers superior perceptual quality across various channel conditions.

% Future works will focus on the lightweight codec architecture. 

% \vspace{-0.1cm}
% \balance
% \def\baselinestretch{1.98}
% \bibliographystyle{IEEEtran} 
% %\def\Baselinestretch{1}
% % \renewcommand{\baselinestretch}{0.8}
% \bibliography{BNN}

\balance

\vspace{-0.1cm}
\bibliographystyle{IEEEtran}
\renewcommand{\baselinestretch}{0.9}
\bibliography{BNN}

\end{document}